\documentclass[aps,prd,twocolumn,preprintnumbers,
eqsecnum,showpacs,nofootinbib,10pt]{revtex4}

\usepackage{amsmath,amsfonts,amssymb,color,graphicx,latexsym,theorem,mathrsfs,comment
}
\usepackage[dvipdfmx]{hyperref}

\newcommand{\ma}[1]{\mbox{$\mathcal{#1}$}}
\newcommand{\mas}[1]{\mbox{$\mathscr{#1}$}}

\newcommand{\D}{{\rm d}}

\newcommand{\ti}{\tilde}
\newcommand{\we}{\wedge}

\begin{document}
%\thispagestyle{empty}

%<<<<<<<<<<<<< TITLE >>>>>>>>>>>>>>>%
\title{
Black holes in an expanding universe and supersymmetry
}

%<<<<<<<<<<<<< AUTHOR >>>>>>>>>>>>>>>%
\author{Dietmar Klemm}
%\email{Dietmar.Klemm_`at'_mi.infn.it}
\author{Masato Nozawa}
%\email{Masato.Nozawa_`at'_mi.infn.it}

%<<<<<<<<<<<<< ADDRESS >>>>>>>>>>>>>>>%

\address{ 
Dipartimento di Fisica, Universit\`a di Milano, and INFN, Sezione di Milano, Via G. Celoria 16, 20133 Milano, Italia
}

%<<<<<<<<<<<<< DATE >>>>>>>>>>>>>>>%
\date{\today}

%======================================%
%<<<<<<<<<<<<< ABSTRACT >>>>>>>>>>>>>>>%
%======================================%
\begin{abstract} 
This paper analyzes the supersymmetric solutions to five and six-dimensional minimal (un)gauged supergravities for which the bilinear Killing vector constructed from the Killing spinor is null. We focus on
the spacetimes which admit an additional ${\rm SO}(1,1)$ boost symmetry. Upon the toroidal dimensional reduction along the Killing vector corresponding to the boost, we show that the solution in the ungauged case describes a charged, nonextremal black hole in a Friedmann-Lema\^itre-Robertson-Walker (FLRW) universe with an expansion driven by a massless scalar field. For the gauged case, the solution corresponds
to a charged, nonextremal black hole embedded conformally into a Kantowski-Sachs universe. It turns out
that these dimensional reductions  break supersymmetry since the bilinear Killing vector and the Killing
vector corresponding to the boost fail to commute.  This represents a new mechanism of supersymmetry
breaking that has not been considered in the literature before.
\end{abstract}

\preprint{IFUM-1045-FT}
\maketitle

%======================================%
%<<<<<<<<<<<<<< SECTION I  >>>>>>>>>>>>>>>>%
%======================================%

\section{Introduction}

Over the last two decades, many developments of superstring theory have been triggered by
supersymmetric solutions in supergravities.  In particular, supersymmetric black holes played a key role for the first successful account for the microscopic origin of the Bekenstein-Hawking entropy~\cite{Strominger:1996sh}. Recently a systematic classification of supersymmetric solutions has been developed and proved useful for obtaining supersymmetric black objects with various topologies
(see e.g.~\cite{Tod:1983pm,Gauntlett:2002nw,Gauntlett:2003fk,Gutowski:2005id,Bellorin:2006yr,
Gutowski:2003rg,Gauntlett:2002fz,Gauntlett:2003wb,Caldarelli:2003pb,Bellorin:2005zc,Meessen:2010fh,
Maeda:2011sh} for an incomplete list).  The supersymmetric solutions are divided into two categories, according to the causal character of the vector field constructed from the Killing spinor, i.e., timelike and
null classes. Typically, the timelike class of solutions contains black holes, whereas the null family contains
propagating waves. The timelike class of metrics in ungauged supergravities is specified by a set of harmonic/Poisson-type functions on a $(d-1)$-dimensional manifold with reduced holonomy over which the metric is fibered.  It therefore follows that supersymmetric black holes belonging to the timelike class are time-independent with degenerate horizons and allow for a superposition principle, as inferred from the Majumdar-Papapetrou solution. This represents a situation in which gravitational and electromagnetic fields are in mechanical equilibrium. 

More than twenty years ago, Kastor and Traschen discovered an interesting generalization of the
Majumdar-Papapetrou solution in the Einstein-Maxwell-$\Lambda (>0)$ system \cite{Kastor:1992nn}. The Kastor-Traschen solution is characterized by a harmonic function on $\mathbb R^3$ with an
additional time-dependence and asymptotically tends to the de Sitter universe. When the harmonic function has a single monopole source at the center of $\mathbb R^3$, the metric describes a black hole with a bifurcate Killing horizon in the de Sitter universe, i.e., the lukewarm limit of the Schwarzschild-de~Sitter
black hole~\cite{Romans:1991nq}. The superposition property of the Kastor-Traschen solution is reminiscent of supersymmetric solutions in the timelike class, although a positive cosmological constant is not
compatible with supersymmetry. Nevertheless, the Kastor-Traschen solution admits a spinor obeying 1st-order differential equations in ``fake'' supergravity, in which the gauge coupling constant in gauged supergravity is analytically continued~\cite{Kastor:1993mj,London:1995ib}. The superposition property further allows to investigate analytically black hole collisions in a (contracting) universe and to test the
validity of the cosmic censorship conjecture~\cite{Brill:1993tm}.  

Later on, ref.~\cite{Maeda:2009zi} obtained a time-dependent and spatially inhomogeneous solution from the time-dependent intersecting M2/M2/M5/M5 branes, which reduces to ${\rm AdS}_2\times S^2$ for $r\to 0$, and approaches for $r\to \infty$ to the FLRW cosmology with the scale factor obeying $a(\tau) \propto \tau^{1/3}$. Maeda and one of the present authors verified that this metric indeed describes a black hole in the FLRW universe with regular horizons~\cite{MN}. The solution was further generalized to the case with a Liouville-type scalar potential, for which the metric asymptotically tends to
an FLRW universe with arbitrary power-law expansion~\cite{Gibbons:2009dr,MNII}. These solutions are very similar to the Kastor-Traschen solution since they are specified by some set of harmonic functions on a base space. Interestingly, the event horizon is generated by an asymptotic Killing vector and realizes the isolated horizon~\cite{Ashtekar:2000sz}, when each harmonic has a point source at the origin. Hence, the area of the horizon fails to grow even though the outside region of the black hole is highly dynamical. Moreover, it
was shown that these solutions are pseudo-supersymmetric in ``fake'' supergravity~\cite{Nozawa:2010zg}. Using the general classification scheme of \cite{Meessen:2009ma}, further extensions to  the case
with a sum of exponential scalar potentials  and to the case including rotation were analyzed
in refs.~\cite{Chimento:2012mg,Chimento:2014afa}.

The cosmic expansion of the solution  in ref.~\cite{Maeda:2009zi} is driven by a massless scalar field corresponding to a ``flat gauging'' in the context of gauged supergravity. It might therefore be possible to embed these solutions into higher-dimensional supersymmetric spacetimes by the Kaluza-Klein
mechanism, rather than embedding them into fake supergravity. As we commented, a naive
Kaluza-Klein reduction does not work, since supersymmetric black holes are time-independent and
extremal, whereas the solution in~\cite{Maeda:2009zi} is time-evolving and non-extremal. To fill this gap is one of the main aims of the present article. 

We exhibit a class of supersymmetric solutions which can be identified as a black hole in an expanding universe upon dimensional reduction. Interestingly, the black hole is time-dependent and admits nondegenerate horizons, both of these properties counter to those for supersymmetric black holes in the timelike class. This is possible because our supersymmetric solutions belong to the null family. We discuss how an additional ${\rm SO}(1,1)$ scaling property gives rise to a Killing vector for the dimensional reduction and how this Kaluza-Klein reduction breaks supersymmetry. This susy breaking mechanism
is new, and may have applications in other contexts as well.

The remainder of our paper is organized as follows. In the next section, we show that the five-dimensional
null BPS family in minimal (un)gauged supergravity admits solutions describing (after a KK reduction)
a black hole in equilibrium
in an expanding universe. In section~\ref{sec:6dim}, we show how to obtain five-dimensional dynamical
black holes from a supersymmetric solution in six-dimensional minimal ungauged supergravity.
Section~\ref{sec:conclusion} contains our conclusions. We employ the mostly plus metric signature
throughout the article.  

\section{Black hole from five dimensions}
\label{sec:5dim}
\subsection{Ungauged case}
\label{subsec:5dim-ungauged}

The bosonic Lagrangian of five-dimensional ungauged minimal supergravity is given by~\cite{Gauntlett:2002nw}
\begin{align}
\label{5Daction}
\ma L_5^{\rm (0)}= R \star 1 -2 F\wedge \star F 
-\frac{8}{3\sqrt 3} F\wedge F \wedge A \,,
\end{align}
where $F=\D A$ is a Maxwell field. In terms of a Dirac spinor $\epsilon$, 
the Killing spinor equation reads
\begin{align}
\label{KSeq_5D}
\hat \nabla_\mu \epsilon\equiv 
\left[\nabla_\mu +\frac{i}{4\sqrt 3}
\left({\gamma_\mu }^{\nu\rho }-4{\delta_\mu }^\nu \gamma^\rho \right)
F_{\nu \rho }\right]\epsilon =0 \,.
\end{align}

Let us consider the case in which $V^\mu \equiv i \bar \epsilon \gamma^\mu \epsilon$
is a null vector. In the coordinate system $V=\partial/\partial v$, the metric and the gauge field are $v$-independent and the general supersymmetric solution in the null family is given by~\cite{Gauntlett:2002nw}
\begin{align}
\label{BPSsol5D}
\D s^2=-2e^+e^-+e^ie^i \,, \qquad A=-\frac{\sqrt 3}2 \ti A_i \D x^i\,,
\end{align}
where $i,j...=1,2,3$ and the orthonormal frame is given by
\[
%\label{5DmetricSUSY}
e^+=H^{-1}\D u \,, \quad e^-=\D v+\frac{\ma F}2 \D u \,, \quad 
e^i=H(\D x^i+a^i \D u)\,.
\]
In three-dimensional vector notation, 
the supersymmetric solutions are determined by the system 
%------------------ dF=0  ----------------%
\begin{align}
& \nabla \times \ti{\bf A} =\nabla H \,, \qquad 
 \partial _u \ti{\bf A} =\frac 13 H^{-2}\mathbf \nabla \times \left(H^3 \mathbf  a \right)\,,
\label{5D_omegaeq}
%\partial_u  \mathbf \nabla H=\frac 13 \mathbf  \nabla \times
% \left[H^{-2}\mathbf \nabla \times \left(H^3 \mathbf  a \right)\right]\,,
\\
&\nabla^2 \ma F=2 H^2 D_u W_{ii} +2 HW_{(ij)}W_{(ij)}
+\frac {2}3HW_{[ij]}W_{[ij]}\,,  \notag 
\end{align}
where 
$D_u \equiv \partial_u -{\bf a}\cdot {\bf \nabla}$ and $W_{ij}\equiv D_u H\delta_{ij}-H\partial_j a_i$.  The integrability condition of (\ref{5D_omegaeq}) leads to 
$\nabla^2 H=0$. The solution to the Killing spinor equation (\ref{KSeq_5D}) is given by the constant spinor under the projection $\gamma^+\epsilon=0$, viz, the solution preserves half of the supersymmetries. 

Let us focus here on the following class of supersymmetric solutions
\begin{align}
\label{aHF}
{\bf a}=0\,, \qquad 
H=H({\bf x})\,, \qquad \ma F= -\frac 4{(hu)^{2}} U({\bf x}) \,. 
\end{align}
With these restrictions, the metric is invariant under the ${\rm SO}(1,1)$ boost action
 $u\to \lambda u$, $v\to v/\lambda$~\cite{Gibbons:2007zu}. Namely there exists an additional 
 Killing vector $\xi=u \partial/\partial u -v\partial/\partial v$ corresponding to the scaling. 
By the following coordinate transformation
$(u,v)\to (t,w)$:
\begin{align}
\label{uvtw}
u=\frac 2h e^{-hw/2} \,, \qquad v=t e^{hw/2} \,,
\end{align}
where $h$ is a constant, the scaling Killing vector is transformed into a coordinate vector,
$\xi=-(2/h)\partial /\partial w$. It therefore follows that the metric (\ref{BPSsol5D}) is independent of
$w$ and reads
\begin{align}
\label{5Dmetric}
\D s^2=H^{-1}\D w[2\D t+(ht+U)\D w]+H^2 \D{\bf x}^2 \,,
\end{align}
where $H$ and $U$ obey Laplace's equations
$\nabla ^2 H=\nabla ^2 U=0$  on $\mathbb R^3$. 
One can then reduce the system down to four dimensions by the Kaluza-Klein ansatz
\begin{align}
\label{4Ddimred}
\D s^2=e^{-2\phi/\sqrt 3}(\D w+2A^{(1)})^2 +e^{\phi/\sqrt 3} g_{\mu\nu} \D x^\mu \D x^\nu \,, 
\end{align}
where 
\begin{align}
\label{}
\phi= \frac{\sqrt 3}2 \ln \left(\frac{H}{ht +U}\right)\,, \qquad 
A^{(1)}= \frac{\D t}{2(ht +U)}\,,
\end{align} 
and the 4-dimensional metric $\D s_4^2=g_{\mu\nu}\D x^\mu \D x^\nu$ reads
\begin{align}
\label{4DBH}
\D s_4^2=- \Xi_4^{-1}  \D t^2+\Xi _4\D {\bf x}^2 \,, 
\end{align}
with $\Xi_4\equiv [(ht+U)H^3]^{1/2}$. This recovers the solution obtained by the compactification of dynamically intersecting branes (with three equal charges)~\cite{Maeda:2009zi}  and solves the four-dimensional field equations derived from the Lagrangian
\begin{align}
\ma L_4^{(0)}=&R-\frac 12(\nabla\phi)^2 \label{4Dredaction}\\ 
& -e^{-\sqrt 3\phi}F^{(1)}_{\mu\nu}F^{(1)\mu\nu}
-e^{-\phi/\sqrt 3}F^{(2)}_{\mu\nu}F^{(2)\mu\nu}\,,
\notag
\end{align}
where $F^{(1,2)}=\D A^{(1,2)}$ and $A^{(2)}=-\frac{\sqrt 3}{2}\ti A_i \D x^i$ descends 
from the five-dimensional gauge potential (\ref{BPSsol5D}). 

Working in spherical coordinates
$\D {\bf x}^2=\D r^2+r^2(\D \theta^2+\sin^2\theta\D \phi^2)$, 
let us consider the case in which only the monopole sources are nonvanishing as 
$H=1+Q/r$ and $U=Q/r$. Asymptotically for $r\to\infty$, the metric 
(\ref{4DBH}) then tends to an expanding FLRW universe
$\D s^2=-\D \tau^2+a^2(\tau)\D {\bf x}^2$,  
where $a\propto \tau^{1/3}$ and $\tau\propto t^{3/4}$. 
%Here and in what follows, we assume $Q>0$ and $h>0$. 
%The metric 
%One notices that for $r\to 0$ with $t $ being finite, the metric can be approximated by 
%${\rm AdS}_2 \times S^2$, which is a typical near-horizon geometry for an extremal black hole. 
%A more detailed analysis~\cite{MN} reveals that the event horizon locates at $r\to 0$, $t\to +\infty$ and $tr$ being finite. This is in accordance with the expectation that the event horizon is an infinite redshift surface with respect to an asymptotic observer and exists outside the trapped region. A simple way to zoom up the vicinity of the horizon is to consider the scaling limit $t\to t/\epsilon$, $r\to \epsilon r$ with $\epsilon\to 0$.  In this limit we have $\Xi_4 \simeq r^{-2}[Q^3(htr+Q)]^{1/2}$, hence by changing to the coordinate 
%\[
%\label{}
%T=\frac{\ln({ht})}h+\int\frac{4h Q^6 R^7\D R}{(R^4-Q^4)f(R)}\,, ~~
%R=[Q^3(h t r+Q)]^{1/4}\,,
%\]
%with 
%\begin{align}
%\label{}
%f(R)&=(R^4-R_+^4)(R^4-R_-^4)\,,\notag \\
%R_\pm ^2&\equiv \frac 12 Q^2[\sqrt{(hQ)^2+4}\pm h Q]\,,
%\end{align}
%the neighborhood of the candidate horizon ($r\to 0$, $t\to +\infty$, $tr$ being finite) is approximated by
%\begin{align}
%\label{}
%\D s_{4,{\rm NH}}^2\simeq 
%-\frac{f(R)}{Q^6R^2}\D T^2+\frac{16R^8\D R^2}{f(R)}
%+R^2 \D \Omega_2^2\,.
%\end{align}
%In this coordinate system, the generator of the horizon is given by $k=\partial/\partial T$.
%One thus sees that  $R_+$ is the black-hole event horizon and the horizon is not degenerate on account of $R_+>R_-$.
As shown in \cite{MN}, the  metric (\ref{4DBH}) then describes a nonextremal black hole in an expanding
FLRW universe for which the cosmic expansion is driven by the massless scalar field. Interestingly, the solution admits a black hole event horizon for which the area is constant even if the outside region of the black hole is highly dynamical. This is a realization of isolated horizons~\cite{Ashtekar:2000sz} and their areal radii are given by $R_\pm ^2\equiv \frac 12 Q^2(\sqrt{(hQ)^2+4}\pm h Q)$~\cite{MN}.  
A similar solution was obtained in \cite{Gibbons:2007zu} by the same scaling method, but it 
fails to admit regular horizons. 

It is worth emphasizing that the bilinear Killing field 
$V=\partial/\partial v=e^{-h w/2} \partial/\partial t $ and 
the Kaluza-Klein Killing field $\xi=u\partial/\partial u-v\partial/\partial v=-(2/h)\partial/\partial w$ do not commute,
\begin{align}
\label{xiV_comm}
[\xi, V]=V \,. 
\end{align}
In other worlds, the Killing spinor is not invariant under the action of the Kaluza-Klein Killing vector. To
see this explicitly, we present the solution to the Killing spinor equation (\ref{KSeq_5D}) in the coordinate
system (\ref{5Dmetric}) for the reader's convenience. 
Introducing the frame
\begin{align}
e^0&=\frac{\D t}{\sqrt{(ht+U)H}} \,, \quad e^i=H \D x^i \,, \notag \\
e^4&=\sqrt{\frac{ht+U}{H}}\left(\D w+\frac{\D t}{ht+U}\right)\,,
\end{align}
and taking the orientation to $\epsilon_{01234}>0$ and $\gamma_{01234}=-i$, 
the solution to (\ref{KSeq_5D}) is given by
\begin{align}
\label{KSsol5D}
\epsilon=[e^{hw}H(ht+U)]^{-1/4}\epsilon_0\,, \qquad 
\gamma_{04}\epsilon_0=\epsilon _0\,,
\end{align}
where $\epsilon_0$ is a constant spinor. 
One sees immediately that $\mas L_\xi \epsilon \ne 0$. This means that the five-dimensional Killing
spinor (\ref{KSsol5D}) does not give rise to a four-dimensional Killing spinor for the solution (\ref{4DBH}),
i.e., the ${\rm U}(1)$ dimensional reduction (\ref{4Ddimred}) breaks supersymmetry. 

One can uplift the BPS solution (\ref{BPSsol5D}), (\ref{aHF}) into eleven-dimensional supergravity, by simply adding a flat torus $T^6$. The resulting solution preserves $1/8$ supersymmetry and describes the intersecting M5/M5/M5 branes (with three equal charges) with a plane wave~\cite{Klebanov:1996mh}. After dimensional reduction to ten dimensional string frame and performing T-duality, one obtains the dynamically intersecting D2/D2/D4/D4 branes. This solution breaks supersymmetry  by the same reasoning as (\ref{xiV_comm}). If this solution is embedded back into eleven dimensions, one obtains the dynamically intersecting M2/M2/M5/M5 branes. A comprehensive analysis of dynamically intersecting branes can be found in~\cite{Maeda:2009zi}. 

Note that the equations of motion derived from (\ref{4Dredaction}) are invariant under 
$F^{(1)}\to \ti F^{(1)} =-e^{-\sqrt 3\phi}\star _4 F^{(1)} $, 
$F^{(2)}\to \ti F^{(2)} =-e^{\phi/\sqrt 3}\star _4 F^{(2)}$ 
and $\phi\to \ti \phi =-\phi$. 
%\begin{align}
%\label{}
%F^{(1)}&\to \ti F^{(1)} =-e^{-\sqrt 3\phi}\star _4 F^{(1)} \,, \notag \\
%F^{(2)}&\to \ti F^{(2)} =-e^{\phi/\sqrt 3}\star _4 F^{(2)} \,, \\
%\phi&\to \ti \phi =-\phi\,.\notag 
%\end{align}
Using these dualized quantities, one can uplift the four-dimensional solution (\ref{4DBH}) back to five dimensions by
$\D s_5^2=e^{-2\ti\phi/\sqrt3}(\D w+2\ti A^{(1)})^2+e^{\ti \phi/\sqrt 3} g_{\mu\nu}\D x^\mu \D x^\nu $ and $\ti F^{(1)}=\D \ti A^{(1)}$. The dualized solution then reads~\cite{Kanou:2014rya}
\begin{align}
\label{Kanou}
\D  s_5^2&=-H^{-2}\D t^2+H 
\D s_{\rm GH}^2 \,,
\qquad 
A= \frac{\sqrt 3}2 \frac{\D t}{H} \,, 
\end{align}
where the base space
$\D s_{\rm GH}^2=h_{mn}\D x^m \D x^n$ is the time-dependent Gibbons-Hawking space~\cite{Gibbons:1979zt},
\begin{align}
\label{}
\D s_{\rm GH}^2=(h t+U)^{-1}(\D w+\chi)^2 +(h t+U)\D {\bf x}^2\,,
\end{align}
with  $\nabla \times \chi =\nabla U$ and $\nabla ^2 U=0$.  
The metric (\ref{Kanou}) is an exact solution to five-dimensional minimal supergravity (\ref{5Daction}) and represents a Kaluza-Klein charged black hole for $H=1+Q/r$ and $U=Q/r$~\cite{Kanou:2014rya}.  In spite of the striking similarity to the canonical form of the metric in the timelike class~\cite{Gauntlett:2002nw}, the metric (\ref{Kanou}) fails to preserve any supersymmetries within the framework of minimal supergravity. This can be checked by computing the integrability condition 
${\rm det}[\hat \nabla_\mu , \hat \nabla_\nu]=0$ for the Killing spinor (\ref{KSeq_5D}). 

The metric (\ref{Kanou})  was originally found by Kanou et al.~in ref.~\cite{Kanou:2014rya}. 
Recently, Ishihara, Kimura and Matsuno pointed out that the metric (\ref{Kanou}) with $H=1+Q/r$,
$U=0$ can be interpreted as a black string in the five-dimensional Kasner universe~\cite{IKM}.

\subsection{Gauged case}

The bosonic Lagrangian of minimal gauged supergravity in five dimensions reads
\begin{align}
\label{5Daction-gauged}
\ma L_5^{(g)} =\ma L_5^{(0)}+12 g^2 \star 1 \,,
%(R+12g^2) \star 1 -2 F\wedge \star F 
%-\frac{8}{3\sqrt 3} F\wedge F \wedge A \,,
\end{align}
where $\ma L_5^{(0)}$ is the ungauged Lagrangian (\ref{5Daction}) and $g\,(>0)$ denotes the gauge coupling constant. The general lightlike supersymmetric solutions to this
theory were classified in~\cite{Gauntlett:2003fk}, and are given by
%\footnote{In this subsection, the five-dimensional geometries are described by the
%coordinates $\{u,v,x^1,x^2,x^3\}$, where $x^1=z$, $x^2=x$ and $x^3=y$.
%Latin letters $i,j,\dots$ are indices on the three-dimensional flat space
%parametrized by $\{x^1,x^2,x^3\}$. Early greek letters $\alpha,\beta,\dots$ are indices
%of the two-dimensional space $\{x^2,x^3\}$, again with flat metric. The antisymmetrc
%tensor $\varepsilon_{\alpha\beta}$ on this space is defined such that
%$\varepsilon_{23}=1$, and $\Delta^{(2)}=\partial_\alpha\partial_\alpha$ is the flat
%Laplacian in two dimensions.} 
\begin{eqnarray}\label{gsol}
\D s^2&=&-H^{-1}(\mathcal{F}\D u^2+2\D u\D v)+H^2\left[(\D x^1+a_1\D u)^2\right. \nonumber \\
&& \left.+e^{3\phi}(\D x^{\alpha} + e^{-3\phi}a_{\alpha} \D u)^2\right]\ ,\nonumber \\
A&=&A_u \D u+\frac{\sqrt3}{4g}\varepsilon_{\alpha\beta}\partial_\alpha \phi \D x^{\beta}\ .
\end{eqnarray}
Here $x^1=z$, $\alpha=2,3$, $x^2=x$, $x^3=y$ and   
$\epsilon_{\alpha\beta}=(i\sigma_2)_{\alpha\beta}$ is an antisymmetric tensor. 
The function $\phi(u,x^i)$ is determined by the equation
\begin{equation}\label{gphi}
e^{2\phi}\partial^2_z e^{\phi}+\Delta^{(2)}\phi=0\,,
\end{equation}
where $\Delta^{(2)}=\partial_\alpha\partial_\alpha$ is a flat Laplacian. 
Given a solution of (\ref{gphi}), $H(u,x^i)$ and 
$A_u(u,x^i)$  are successively obtained  from 
\begin{align}
&H=-\frac{1}{2g}\phi' \,, \\ 
\label{gAu}
&[H^2 e^{2\phi}(e^\phi A_u)']'+\partial_\alpha (H^2\partial_\alpha A_{u})=
\frac{\sqrt3}{2g} H 
\varepsilon_{\alpha \beta} \partial_\alpha \dot \phi \partial_\beta H\,.
\end{align}
Dots and primes denote respectively the derivatives with respect to $u$ and $z$. 
Then the functions $a_i(u,x^j)$ are determined by the system
\begin{align}
\label{ga}
&\qquad \quad ~
\frac 1{2\sqrt3}\varepsilon_{\alpha\beta}\partial_{\alpha}(H^3 a_\beta)=-H^2 e^{2\phi}
\partial_z(e^\phi A_u)\,, \\
&\frac 1{2\sqrt3}[\partial_\alpha(H^3 a_1)-(H^3a_\alpha)']= H^2
\varepsilon_{\alpha\beta} \partial_\beta A_{u} -\frac{\sqrt3}{4g} H^2 \partial_\alpha\dot \phi\ , \nonumber
\end{align}
whose integrability condition is (\ref{gAu}). Finally, the function $\mathcal{F}(u,x^i)$
follows from the $uu$-component of the Einstein equations derived from (\ref{5Daction-gauged}). 
The solution preserves $1/4$ of the supersymmetries. 

Since in the five-dimensional ungauged case, the metric (\ref{5Dmetric}) describes a wave on a black
string, we would like to obtain a similar BPS solution in the gauged theory as well. To this end,
we follow the construction in \cite{Bernamonti:2007bu}, and suppose $\phi$ to be separable,
\begin{equation}
\phi(u,x,y,z)=\phi_1(z)+\phi_2(x,y)\ .
\end{equation}
Substituting this expression of $\phi$ into (\ref{gphi}) we find that $\phi_1$
and $\phi_2$ have to satisfy the equations
\begin{eqnarray}
 \label{phi1}\partial^2_z e^{\phi_1}&=&\frac{k}{24g}e^{-2\phi_1}\ ,\\
\label{phi2}\Delta^{(2)}\phi_{2}&=&-\frac{k}{24g}e^{3\phi_2}\ ,
\end{eqnarray}
where $k$ is a constant. (\ref{phi1}) implies
\begin{equation}
 e^{3\phi_1} (\phi_{1}')^2=\mu e^{\phi_1}-\frac{k}{12g}\ ,
\end{equation}
where $\mu$ denotes another integration constant. 
As a particular solution of the Liouville equation (\ref{phi2}), we choose
\begin{equation}
 e^{3\phi_2}=64g\Upsilon^{-2}\ ,
\end{equation}
where $\Upsilon(x,y)=1+k(x^2+y^2)$. With these choices, the system \eqref{ga} is satisfied for
$A_u=a_1=a_\alpha=0$. If we introduce the new radial coordinate
\begin{equation}
\rho = \frac1{2\phi_{1}'(g e^{\phi_1})^{3/2}}\,,
\end{equation}
the metric and gauge field \eqref{gsol}  become
\begin{align}
\D s^2 &= \frac{1}{(g\rho)^2}\left[-f^{3/2}\left(\frac{\ma F}{2}\D u^2+\D u \D v\right)
+\frac{\D \rho^2}{f^2}+g^{-2}\D \Sigma_k^2 
\right]\,,
%\frac{f^{3/2}}{(g\rho)^2}\!\left({\cal F}\D u^2 +2\D u\D v\right) +
%\frac{\D\rho^2}{(g\rho f)^2}+(g^2\rho)^{-2} \D \Sigma_k^2\,, 
\nonumber \\
A &=\frac k{g\sqrt3\Upsilon}(y\D x - x\D y)\ , \label{wave-on-string-AdS}
\end{align}
where we defined $f=1+g^2k\rho^2/3$ and the constant $\mu$ has been eliminated by a rescaling of
$u,v$. $\D \Sigma_k^2=4\Upsilon^{-2}(\D x^2+\D y^2)$ is the line element of the unit constant
curvature space with $k=0,\pm 1$. 
\eqref{wave-on-string-AdS} represents a wave on
a magnetic string in $\text{AdS}_5$, with wave profile $\cal F$ determined by the $uu$-component of
the Einstein equations, which in the present case boils down to
\begin{equation}
\Upsilon^2\Delta^{(2)}{\cal F} + \frac{8f^2 - 20f}{g^2\rho}\partial_\rho{\cal F} +
\frac{4f^2}{g^2}\partial^2_\rho{\cal F} = 0\ . \label{equ-profile}
\end{equation}
Similar to the ungauged case, we seek for a solution of the form ${\cal F}=-4(hu)^{-2}U(\rho)$,
with $h$ a constant. Then, \eqref{equ-profile} can be easily solved, with the result
\begin{equation}
U = C_1\frac{2 + g^2 k\rho^2}{f^{3/2}} + C_2\,, 
\label{Usol}
\end{equation}
where $C_{1,2}$ denote integration constants. Notice that, in the limit $\rho\to 0$, the
solution \eqref{wave-on-string-AdS} asymptotes to (magnetic) $\text{AdS}_5$. If $U=0$,
the horizon is located at the zeroes of $f$, so one has a genuine black string in the hyperbolic case
$k<0$ (constructed in \cite{Klemm:2000nj}), a naked singularity for $k>0$ (found
in \cite{Chamseddine:1999xk}), and $\text{AdS}_5$ for $k=0$.

The metric \eqref{wave-on-string-AdS} is again invariant under the scaling
$u\to\lambda u$, $v\to v/\lambda$, so we can follow the same procedure as in subsection
\ref{subsec:5dim-ungauged}, namely introduce the new coordinates $t$ and $w$ by
(\ref{uvtw}), and then Kaluza-Klein reduce along the Killing direction $\partial/\partial w$, using the ansatz \eqref{4Ddimred}.
This leads to the four-dimensional solution
\begin{equation}
\D s_4^2 = -\Xi_4^{-1}\D t^2 + \frac{\Xi_4 f^3}{4(g\rho)^6}\left(\frac{\D\rho^2}{f^2} +
g^{-2}\D \Sigma_k^2 \right)\ , \label{metr-BH-Kant-Sachs}
\end{equation}
where $\Xi_4 \equiv{2\sqrt2(g\rho)^3}{f^{-9/4}}\left(ht + U\right)^{1/2}$. 
The dilaton and Kaluza-Klein gauge field are given by
\begin{equation}
e^{-2\phi/\sqrt3} = \frac{f^{3/2}}{2g^2\rho^2}(ht + U)\ , \quad A^{(1)} = \frac{\D t}{2(ht + U)}\ .
\label{phi-A-BH-Kant-Sachs}
\end{equation}
\eqref{metr-BH-Kant-Sachs} and \eqref{phi-A-BH-Kant-Sachs} solve the equations of motion derived
from the four-dimensional Lagrangian
\begin{align}
\label{4Dredaction:gauged}
\ma L_4^{(g)}=
\ma L_4^{(0)}+12 g^2 e^{\phi /\sqrt 3} \,, 
%&R-\frac 12(\nabla\phi)^2  + 12g^2 e^{\phi/\sqrt3}\notag\\
%& -e^{-\sqrt 3\phi}F^{(1)}_{\mu\nu}F^{(1)\mu\nu}
%-e^{-\phi/\sqrt 3}F^{(2)}_{\mu\nu}F^{(2)\mu\nu}\,,
\end{align}
where $\ma L_4^{(0)}$ was defined by (\ref{4Dredaction}). 
For the solution considered here, $A^{(2)}$ is given by \eqref{wave-on-string-AdS}.
\eqref{4Dredaction:gauged} represents (after dualization of e.g.~$F^{(1)}$)
the zero-axion truncation of the $\text{t}^3$ model of $N=2$ supergravity with
Fayet-Iliopoulos gauging, leading to the Liouville potential for $\phi$.

Notice that $U$ defined in \eqref{Usol} is still a harmonic function, but now on the curved base space
\begin{equation}
h_{mn}\D x^m\D x^n = \frac{f^3}{4(g\rho)^6}\left(\frac{\D\rho^2}{f^2} +
g^{-2}\D \Sigma_k^2\right)\,.
\end{equation}
Let us move to the physical discussion for the solution (\ref{metr-BH-Kant-Sachs}). 
Due to the freedom $t\to t+t_0$, one can choose $2C_1 + C_2=0$ in (\ref{Usol}), for which 
$U(\rho=0)=0$. 
Since 
\begin{align}
\label{}
\int _Se^{-\sqrt 3\phi} \star F^{(1)}=\frac{k^2\Sigma_kC_1}{3g} \,, \quad
\int_S  F^{(2)}=\frac{k\Sigma_k}{\sqrt 3g}\,, 
\end{align}
the magnetic charge obeys a Dirac-type quantization condition and  the electric charge is proportional
to $C_1$ which we shall assume to be negative in what follows. 
In the asymptotic region $\rho \to 0$, 
the metric \eqref{metr-BH-Kant-Sachs} reduces to
\begin{equation}
\D s_4^2 = \frac1{2\sqrt2(g\rho)^3}\left[-\D\tau^2 + a^2(\tau)\left(\D\rho^2 +
g^{-2}\D \Sigma_k^2\right)\right]\ , \nonumber
\end{equation}
where $\tau \propto t^{3/4}$ and $a(\tau ) \propto (h\tau)^{1/3}$. 
The asymptotic geometry is thus conformal to a Kantowski-Sachs universe with power-law
expansion for $h>0$. Note that in a generic Kantowski-Sachs universe, the part proportional to $\D x^2 + \D y^2$ and $\D\rho^2$ can have different scale factors. Here they happen to be equal. 
The behavior of the scale factor $a(\tau ) \propto (h\tau)^{1/3}$ is the same as that driven by a massless scalar in an FLRW universe. However, this does not mean that the potential plays no role, since 
the metric depends also on the coordinate $\rho$. 

Since the five-dimensional $U=0$ black string has a horizon for $k<0$, we shall focus on this case in
what follows. In terms of $r=1/(g^2\rho)$, $f=0$ has a root at $r=r_+\equiv (\sqrt 3 g)^{-1}$. This is a
naive horizon locus for the metric (\ref{metr-BH-Kant-Sachs}). As pointed out in \cite{MN}, we have to
take the $t\to\infty $ limit at the same time, since the event horizon is an infinite redshift surface. To see
the geometry of this candidate horizon, let us take the near-horizon limit 
\begin{align}
\label{scaling}
t=\frac{\hat t}{\epsilon^{3/2}}\,, \qquad r-r_+ =\epsilon \hat r \,, \qquad \epsilon\to 0\,.
\end{align}
The near-horizon metric is independent of $\epsilon$ and reads
\begin{align}
\label{NHmetric_tr}
\D s^2 _{\rm NH}= -\frac{2 g\hat r^3 }{3\sqrt 3 R^2 } \D \hat t^2+\frac{3R^2 \D \hat r^2}{4 \hat r^2}+R^2 \D \Sigma_{k=-1}^2 \,, 
\end{align}
where $(gR)^2=2^{-3/4} 3^{-13/8}\left(
12 h \hat t (g\hat r)^{3/2}-2^{1/2} 3^{1/4} C_1 
\right)^{1/2}$. 
As a consequence of the scaling limit (\ref{scaling}), 
there appears an asymptotic Killing vector
\begin{align}
\label{asy_Killing}
K= \hat r \frac{\partial }{\partial \hat r} - \frac{3}{2}\hat t \frac{\partial}{\partial \hat t}\,, \qquad 
\mas L_K g_{\rm NH}=0 \,. 
\end{align}
Changing coordinates from $(\hat t, \hat r)$ to $(T,R)$, where $T$ is defined by 
\begin{align}
\label{}
 T=\log (\hat r)- \int \frac{144 g^4 R^3(C_1 +54 g^4 R^4)}{ f_1( R)} \D  R\,, 
\end{align}
with $f_1( R) \equiv   (54 g^4R^4 )^2+(108C_1g^2-36h^2) g^2R^4+C_1^2$, 
the near-horizon metric (\ref{NHmetric_tr}) is cast into
\begin{align}
\label{}
\D s_{\rm NH}^2=-\frac{ f_1( R)}{48 g^2 h^2 R^2}\D  T^2 
+\frac{12(36 g^4R^4)^2 }{ f_1( R)}\D  R^2+ R^2 \D \Sigma_{k=-1}^2 \,.
\end{align}
In this coordinate system, the asymptotic Killing vector (\ref{asy_Killing}) 
becomes $K=\partial/\partial T$. 
Therefore, there exist Killing horizons $R=R_\pm$ at roots of $f_1(R)=0$, i.e.,
\begin{align}
\label{}
R_\pm^2 =\frac{g^{-2}}{18 }\left(\pm hg^{-1}+\sqrt{h^2g^{-2}-6C_1}\right) \,. 
\end{align}
This makes it obvious that the horizon is nonextremal. 

\begin{figure}[t]
\begin{center}
\includegraphics[width=4cm]{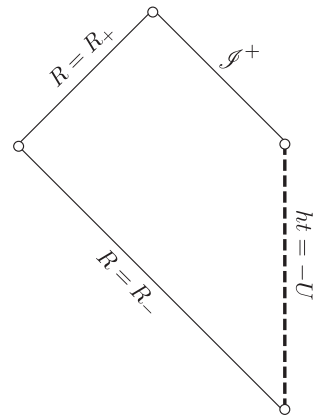}
\caption{Penrose diagram of the black hole embedded in the conformal Kantowski-Sachs universe.
There are regular horizons at $R=R_\pm$.}
\label{fig:PD}
\end{center}
\end{figure}

To lend a further credence to the above picture, 
we have traced numerically the radial null geodesic motions and arrived at the 
conformal diagram shown in Fig.~\ref{fig:PD}. The causal nature is analogous to that in the ungauged solution (\ref{4DBH}) (see the conformal diagram in~\cite{MN}), but the asymptotic structure is quite different.  There appears a timelike naked singularity at $ht+U=0$, but otherwise the metric behaves non-pathologically. As is clear from Fig.~\ref{fig:PD}, the spacetime (\ref{metr-BH-Kant-Sachs}) admits a 
regular, nonextremal event horizon at $R=R_+(>R_-)$, which remains constant in time. The appearance of these isolated horizons is ascribed to the near-horizon asymptotic Killing vector (\ref{asy_Killing}). 

Note finally that for $k=-1$, the scalar field $\phi$ and gauge fields $A^{(i)}_\mu \D x^\mu$ also admit a
definite limit under (\ref{scaling}), and the resulting system (\ref{NHmetric_tr}) solves the same field
equations  derived from (\ref{4Dredaction:gauged}) as the original solution. This means that the scaling
limit (\ref{scaling}) is indeed well-defined.

\section{Black hole from six dimensions}
\label{sec:6dim}

Let us next consider minimal ungauged supergravity in six dimensions~\cite{Gutowski:2003rg}. 
The equations of motions are given by
%------------  Einstein equations   --------------%
\begin{align}
R _{\mu\nu }=G_{\mu \rho\sigma }G_{\nu }{}^{\rho \sigma }
 \,, \qquad \D G=0 \,, \qquad G =-\star G \,, 
\end{align}
where $G$ is the three-form field strength. Since $G$ is anti-self-dual, 
there exists no covariant action which gives rise to the above equations. 
The Killing spinor equation reads  
%------------  Killing spinor  --------------%
\begin{align}
\label{6DKS}
\hat \nabla_\mu \epsilon \equiv\left(\nabla_\mu -\frac{1}{4}G_{\mu\nu\rho }
\gamma^{\nu\rho }\right)\epsilon =0\,, 
\end{align}
where $\epsilon$ is an anti-chiral spinor, $\gamma_7 \epsilon =-\epsilon$,
with $\gamma_7=\gamma_{012345}$. 

According to the general analysis given in \cite{Gutowski:2003rg}, there appears only the null family, and the general supersymmetric metric can be written as 
$\D s^2=-2 e^+e^-+\delta_{IJ}e^I e^J$ ($I=1,...,4$) with the orientation $\epsilon_{+-1234}>0$, where
\begin{align}
e^-&=-\left(\D v+\omega-\frac{\ma FH}{2}e^+\right)\,, \notag\\
e^+&=-H^{-1}(\D u+\beta)\,, \qquad 
e^I=H^{1/2} \hat e^I \,,
\end{align}
and 
%--------------------  Field strength   ------------------------%
\begin{align}
2G=&\star _4 (\ma DH+H\dot \beta )+(H^{-1}\ma DH+\dot \beta )
\we e^+\we e^-\notag \\&
+[H\ma G^--(\ma D\omega )^-]\we e^+ +H^{-1}e^-\we \ma D \beta \,. 
\end{align}
Here the metric and the three-form field are $v$-independent. $h_{mn}=\delta_{IJ}\hat e^I{}_m \hat e^J{}_n$
is the base-space metric of an integrable almost hyper-K\"ahler manifold satisfying 
\begin{align}
 \hat \D J^i=\partial_u (\beta\we J^i)\,, 
\label{min_Jeq}
\end{align}
where $\hat{\D}$ is the exact differential on the base space and the $J^i$ satisfy
$J^i\cdot J^j=-\delta^{ij}+\epsilon_{ijk}J^k$. 
$\beta$ and $\omega$ are one-forms on the base space with a $u$-dependence
obeying $\ma D \beta =\star _4 \ma D \beta$, where $\ma D$ is a linear operator 
acting on $p$-form fields on the base space as $\ma D\equiv \hat \D -\beta \we \partial_u$.  
The supersymmetric system is determined by solving 
\begin{align}
 \ma D [\star_4 (\ma D H+H \dot \beta )]-\ma D\beta\we \ma G^+&= 0\,,\label{minimal_Heq}\\
 \hat \D \ma G-\partial_u [\beta\we \ma G-\star_4(\ma DH+H\dot \beta )]&=0\,,
%\label{minimal_Geq}
\notag
\end{align}
and 
\begin{align}
- \star _4 \ma D(\star_4 L)=& \frac{1}{2}H h^{mn} \partial_u^2(H h_{mn})+\frac{1}{4}
\partial_u (Hh^{mn})\partial_u (H h_{mn})
\nonumber \\
&-2\dot \beta_m L^m +\frac{1}{2}H^{-2}[(\ma D \omega )^--H \ma G^-]^2
\notag \\
& -\frac{1}{2}H^{-2}
\left(\ma D \omega +\frac{1}{2}\ma F \ma D \beta \right)^2 \,,
\label{minimal_Leq}
\end{align}
where
$\ma G=\ma G^++\ma G^-$, $\star_4\ma G^\pm =\pm \ma G^\pm$ 
and $L$ are given by 
\begin{align}
\label{}
\ma G^+&\equiv H^{-1}\left[(\ma D \omega )^++\frac{1}{2}\ma F \ma D \beta \right]\,,\notag \\
\ma G^-&\equiv \frac 18 H\epsilon_{ijk}(J^i)^{pq}(\dot J^j)_{pq} J^{k} \,, \\
 L&\equiv \dot \omega+\frac{1}{2}\ma F \dot\beta -\frac{1}{2}\ma D \ma F\,,\notag 
\end{align}
where the dot denotes a differentiation with respect to $u$.
Under the projection $\gamma^+\epsilon=0$, the solution to the Killing spinor equation (\ref{6DKS})
is given by a constant spinor. 

Let us consider the case where
\begin{align}
\label{}
\beta=0\,, \quad \dot H=\dot h_{mn}=0\,, \quad 
(\hat \D \omega)^+=0\,.
\end{align}
Eq.~(\ref{minimal_Heq}) gives $\Delta_h H=0$, while (\ref{minimal_Leq}) leads to
$\Delta_h \ma F=0$ provided we work in the gauge $\hat \D \star_4 \omega=0$.  
As in the previous section, we choose 
\begin{align}
\label{}
\omega =\frac{2}{hu} \varpi \,, \qquad 
\ma F=-\frac{4}{(hu)^2}U\,, 
\end{align}
where $U$ and $\varpi$ are a $u$-independent scalar and one-form respectively, 
and consider the coordinate transformation (\ref{uvtw}). One sees that the solution is independent of $w$ and the spacetime can be dimensionally reduced as 
$\D s_6^2=e^{-\sqrt{3/2}\phi}(\D w+2A^{(1)})^2+e^{\phi/\sqrt 6}\D s_5^2$, 
where 
\begin{align}
\label{5DsolphiA}
\phi =\sqrt{\frac 23}\log\left(\frac{H}{ht +U}\right)\,, \quad A^{(1)}=
\frac 12 \frac{\D t+\varpi}{(ht+U)}\,.
\end{align}
The five-dimensional metric becomes the one found in~\cite{Klemm:2000gh,Behrndt:2003cx},
\begin{align}
\label{5Dsol}
\D s^2_5=-\Xi_5 ^{-2} (\D t+\varpi)^2 +\Xi_5 h_{mn}\D x^m \D x^n  \,, 
\end{align}
where $\Xi_5\equiv [(ht+U)H^2]^{1/3}$ and $\hat \D \omega$ is an anti-self-dual two-form on the
hyper-K\"ahler base space $h_{mn}$. $H$ and $U$ satisfy $\Delta_hH=\Delta_h U=0$. 
The solution (\ref{5Dsol}) solves the field equations derived from the action
\begin{align}
\label{effaction5D}
\ma L_5=&R-\frac 12 (\nabla\phi)^2 -e^{-2\sqrt{2/3}\phi}F^{(1)}_{\mu\nu}F^{(1)\mu\nu}\\
 &-e^{\sqrt{2/3}\phi}F^{(2)}_{\mu\nu}F^{(2)\mu\nu}
 +\epsilon^{\mu\nu\rho\sigma\tau} A^{(1)}_{\mu} F ^{(2)}_{\nu\rho} F^{(2)}_{\sigma\tau} \,,
 \notag
\end{align}
where $A^{(2)}=\frac 1{\sqrt 2}H^{-1}(\D t+\varpi)$. 

Taking the Euclidean space 
$\D r^2+r^2 (\D \vartheta^2+\cos^2\vartheta \D \phi_1^2+\sin^2\vartheta\D \phi_2^2)$ 
as the hyper-K\"ahler base, and 
considering only the lowest-order harmonic contributions $H=1+Q/r^2$, $G=Q/r^2$, 
$\varpi=(j/r^2)(\cos^2\vartheta\D \phi_1+\sin^2\vartheta\D\phi_2)$, 
this solution describes a five-dimensional rotating black hole in an expanding FLRW universe in five dimenisons~\cite{Nozawa:2010zg}. This proves that the five-dimensional dynamical metric (\ref{5Dsol}) admitting regular nonextremal horizons and an ergoregion is supersymmetric from a six-dimensional point
of view. Note that, as in the previous section, the dimensional reduction along
$\xi=-(2/h)\partial/\partial w$ breaks supersymmetry.

\section{Concluding remarks}
\label{sec:conclusion}

In this letter, we pointed out that the null family of supersymmetric solutions in (un)gauged supergravities admits an interesting class of dynamical spacetimes that describe black holes in an expanding universe
upon dimensional reduction. The black hole metrics are dynamical and nonextremal, both of which
are in marked contrast to the properties of conventional supersymmetric black holes belonging to the timelike class. The most interesting aspect of our findings is that the simple toroidal reduction breaks supersymmetry. As far as the authors know, this provides a new supersymmetry-breaking mechanism following from the noncommutativity of the Kaluza-Klein Killing field and the bilinear Killing field
constructed from the Killing spinor. Our results may be useful for generalizing the BPS
attractors~\cite{Hristov:2014eba} into time-dependent settings. 

For the four-dimensional example described in section \ref{sec:5dim}, we have considered only the solution without rotation. One may expect that the rotating solution specified by three harmonic
functions \cite{Chimento:2014afa} could be obtained in a similar fashion.  
The only way to realize this is to set ${\bf a}=-2(hu)^{-1} H^{-3}{\boldsymbol\omega}$ in (\ref{aHF}),
where $\partial_u {\boldsymbol\omega}=0$. From the first two equations in (\ref{5D_omegaeq}), ${\boldsymbol\omega}$ obeys 
$\nabla\times{\boldsymbol\omega}=H\nabla K-K\nabla H$
in terms of another harmonic function $K$. After replacing 
$U\to U-H^{-3}|{\boldsymbol\omega}|^2$ in (\ref{aHF}), one finds that 
$U$ satisfies
\begin{align}
\label{}
\nabla^2U=h \nabla\cdot {\boldsymbol\omega}+\frac 2{3H^3}(K\nabla H-H\nabla K)^2 \,. 
\end{align}
This implies that $U$ obeys the Laplace equation iff $K\propto H$, recovering ${\boldsymbol\omega}=0$.  Therefore, the rotating solution in \cite{Chimento:2014afa} is not obtained from an ${\rm SO}(1,1)$
boost-invariant form. This situation is similar to the stationary counterpart, for which 
the Majumdar-Papapetrou solution can be embedded both into  the timelike and null classes, whereas
the IWP family~\cite{Perjes:1971gv,Israel:1972vx} can only be embedded into the timelike class. 

One can nevertheless consider the monopole harmonics $H=1+Q_1/r$, $K=1+Q_2/r$ in the 
rotating case, for which one gets
\begin{align}
\label{}
U&=k_1+\frac{k_2}r +\frac{(Q_1-Q_2)^2}{3r(r+Q_1)} \,, 
%\notag \\{\boldsymbol \omega}\cdot\D {\bf x} &
\quad \omega =(Q_1-Q_2)\cos\theta \D \phi \,,
\end{align}
where $k_{1,2}$ are constants. This solution, however, always suffers from a
Dirac-Misner string unless $Q_1=Q_2$. Since the metric is $t$-dependent, this singularity cannot be removed by periodic identification of $t$. In conclusion, it remains to be seen if the four-dimensional
rotating dynamical black holes of \cite{Chimento:2014afa} can be embedded in some way into
supersymmetric higher-dimensional spacetimes.

We have thus far discussed the Kaluza-Klein type dimensional reduction, which generates only a massless
field in the ungauged case. One may thus wonder if an analogous higher-dimensional embedding works for generalized dimensional reductions. To illustrate this, let us consider the non-twisting
fake-supersymmetric black holes in a five-dimensional FLRW universe obtained in \cite{Nozawa:2010zg}
and see if they arise from a generalized dimensional reduction of six-dimensional supersymmetric solutions. 
The metric takes the form 
$\D s_5^2=-\Xi_5^{-2} \D t^2+\Xi h^{(\rm HK)}_{mn}\D x^m \D x^n$ with 
$\Xi_5\equiv [(ht+U)^{n}H^{3-n}]^{1/3}$, where the base is a time-independent hyper-K\"ahler
manifold, and the gauge and scalar fields are given by 
\begin{align}
\label{5Dsolpot}
\phi&= \sqrt{\frac{n(3-n)}{3}}\log \left(\frac{H}{ht+U}\right)\,, \notag \\
A^{(1)}&=\frac{\sqrt n}{2 (ht+U)}\D t \,, \qquad A^{(2)}=\frac{\sqrt{3-n}}{2H}\D t \,.
\end{align}
Here $\Delta_h H=\Delta_h U=0$ as before. 
This solution solves the field equations derived from the action
\begin{align}
\label{5Deffactionpot}
\ma L_5=&R-\frac 12 (\nabla\phi)^2 -V(\phi)\notag \\ & 
-e^{-\alpha \phi} F_{\mu\nu}^{(1)}F^{(1)\mu\nu} 
-e^{4\phi/3\alpha}F_{\mu\nu}^{(2)}F^{(2)\mu\nu}\,,
\end{align}
where $V=\frac 12 n(n-1)h^2 e^{\alpha \phi}$ and $\alpha =2\sqrt{(3-n)/(3n)}. $
For $n=1$, the solution (\ref{5DsolphiA})--(\ref{5Dsol}) with $\varpi=0$ is recovered. 
A plausible dimensional reduction to generate this kind of scalar potential is the Scherk-Schwarz 
mechanism~\cite{Scherk:1978ta}, exploiting a global symmetry and imposing twisted boundary conditions. Following the argument in ref.~\cite{Kerimo:2004qx}, one can embed the five-dimensional theory (\ref{5Deffactionpot}) with $n=2$ into six-dimensional minimal {\it ungauged} supergravity coupled to a vector multiplet, whose bosonic action reads
\begin{align}
\label{SS_6Daction}
\ma L_6 =& R-\frac 12 (\nabla\Phi)^2 -\frac 1{12}e^{-\sqrt 2\Phi} 
G_{\mu\nu\rho}G^{\mu\nu\rho}\notag \\& -e^{-\Phi/\sqrt 2} F^{(1)}_{\mu\nu}
F^{(1)\mu\nu} \,. 
\end{align}
The embedding of the solution (\ref{5Dsolpot}) is achieved by the explicit $z$-dependent form 
\begin{align}
\label{SS_6Dsol}
&\D s_6^2=e^{hz/2}\left(e^{\phi/(2\sqrt 6)} \D s_5^2 +e^{-{\sqrt 3\phi}/(2\sqrt 2)}
\D z^2 \right) \,,  \\
&G= \star _h \D H \,, \quad 
F^{(1)}=\D A^{(1)} \,, \quad \Phi=\frac {\sqrt 3 }2\phi-\frac{hz}{\sqrt 2} \,. \notag 
\end{align}
The supersymmetric solutions for the system (\ref{SS_6Daction}) were classified in \cite{Cariglia:2004kk}. 
One can check that the uplifted solution (\ref{SS_6Dsol}) fails to satisfy the integrability condition for the Killing spinor, e.g, ${\rm det}(F^{(1)}_{\mu\nu} \gamma^{\mu\nu})=0$ is not satisfied. Namely, (\ref{SS_6Dsol}) does not preserve any supersymmetry in the theory (\ref{SS_6Daction}). 
It would be interesting to see if the dynamical solutions with scalar potential constructed
in \cite{Nozawa:2010zg} (for $n\neq 1,2$) and in \cite{Gibbons:2009dr,Chimento:2012mg}
can be embedded into BPS solutions in higher dimensions by the Scherk-Schwarz mechanism or by a sort
of brane world reduction \cite{Park:2001jh}.
This would be useful for further elucidating the
Kaluza-Klein network originating from M-theory.

\section*{Acknowledgements}

The impetus of the present work stemmed from the talk by Masashi Kimura at the workshop ``One Hundred Years of Strong Gravity'' held at  Instituto Superior T\'ecnico (IST) in Lisbon. MN is thankful to CENTRA for their kind hospitality. This work is supported partly by JSPS and INFN.

%======================================%
%<<<<<<<<<<<< REFERENCES  >>>>>>>>>>>>>%
%======================================%

\end{document}